\documentclass[epj]{webofc}
\usepackage[utf8]{inputenc}
\usepackage[varg]{txfonts}   % Web of Conferences font
\usepackage{booktabs}
\usepackage{xcolor}
\definecolor{darkred}{rgb}{0.4,0.0,0.0}
\definecolor{darkgreen}{rgb}{0.0,0.4,0.0}
\definecolor{darkblue}{rgb}{0.0,0.0,0.4}
\usepackage[bookmarks,linktocpage,colorlinks,
    linkcolor = darkred,
    urlcolor  = darkblue,
    citecolor = darkgreen]{hyperref}
\usepackage{subfigure}
\usepackage{xspace}
\wocname{EPJ Web of Conferences}
\woctitle{Lattice2017}
\newcommand{\software}[1]{\texttt{#1}\xspace}

\newcommand{\METAQ}{\software{METAQ}}
\newcommand{\SLURM}{\software{SLURM}}
\newcommand{\MOAB}{\software{MOAB}}
\newcommand{\PBS}{\software{PBS}}
\newcommand{\TORQUE}{\software{TORQUE}}
\newcommand{\mpijm}{\software{mpi\_jm}}
\newcommand{\wraprun}{\software{wraprun}}
\newcommand{\bash}{\software{bash}}

\newcommand{\MPI}{\software{MPI}}
\newcommand{\python}{\software{python}}
\newcommand{\QMP}{\software{QMP}}
\newcommand{\YAML}{\software{YAML}}

\begin{document}
%%%%%%%%%%%%%%%%%%%%%%%%%%%%%%%%%%%%%%%%%%%%%%%%%%%%%%%%%%%%%%%%%%%%%%%%%%%%%
%
\selectlanguage{english}
%----------------------------------------------------------------------------
\title{%
Job Management and Task Bundling
}
%----------------------------------------------------------------------------
\author{%
\firstname{Evan}        \lastname{Berkowitz}   \inst{1}\fnsep\thanks{\email{e.berkowitz@fz-juelich.de}.\\  Corresponding slides are available at \url{https://makondo.ugr.es/event/0/session/102/contribution/335}.} \and
\firstname{Gustav~R.}      \lastname{Jansen}       \inst{2}\fnsep
\thanks{This manuscript has been authored by UT-Battelle, LLC under
Contract No. DE-AC05-00OR22725 with the U.S. Department of
Energy. The United States Government retains and the publisher, by
accepting the article for publication, acknowledges that the United
States Government retains a non-exclusive, paid-up, irrevocable,
world-wide license to publish or reproduce the published form of
this manuscript, or allow others to do so, for United States
Government purposes. The Department of Energy will provide public
access to these results of federally sponsored research in
accordance with the DOE Public Access
Plan. (http://energy.gov/downloads/doe-public-access-plan).}
\and
\firstname{Kenneth}     \lastname{McElvain}     \inst{3} \and
\firstname{Andr\'{e}}   \lastname{Walker-Loud}  \inst{3} 
}
%----------------------------------------------------------------------------
\institute{%
Institut f\"{u}r Kernphysik and Institute for Advanced Simulation, Forschungszentrum J\"{u}lich
\and
National Center for Computational Sciences and Physics Division, Oak Ridge National Laboratory
\and
Nuclear Science Division, Lawrence Berkeley National Laboratory
}
%----------------------------------------------------------------------------
\abstract{%
High Performance Computing is often performed on scarce and shared computing resources. To ensure computers are used to their full capacity, administrators often incentivize large workloads that are not possible on smaller systems.
Measurements in Lattice QCD frequently do not scale to machine-size workloads. By bundling tasks together we can create large jobs suitable for gigantic partitions.
We discuss \METAQ and \mpijm, software developed to dynamically group computational tasks together, that can intelligently backfill to consume idle time without substantial changes to users' current workflows or executables.
}
%----------------------------------------------------------------------------
\maketitle
%----------------------------------------------------------------------------
\section{Introduction}\label{intro}

Many large scientific calculations require enormous computational resources that are not available except at leadership-class computing facilities.
To ensure that the computers at these facilities are used to their full potential, administrators often incentivize users to aim for large jobs that cannot be executed except on these large machines.
For example, NERSC discounts the spent computer time by 40\% once jobs are larger than 683 nodes\cite{NERSC.charge} and the queue on Titan, at Oak Ridge National Laboratory, artificially ages jobs larger than 3750 nodes so that they make it through the queue more quickly\cite{OLCF.charge}.

Lattice QCD (LQCD) calculations are not often suited to such large jobs.
For example, the calculation of the nucleon axial coupling $g_A$ presented by Chang at LATTICE 2017\cite{jason,Berkowitz:2017gql} required thousands of propagators, thousands of sequential (Feynman-Hellman) propagators\cite{Bouchard:2016heu}, and thousands of contractions.
In that work the authors typically solved for the propagator using 32 nodes, each with a GPU, while the contractions were performed on 8 nodes using only CPUs.
To run any of these tasks on hundreds or even thousands of nodes would be a waste of resources as the data would be spread too thin. 
Consequently, the CPUs and GPUs would largely sit idle while communication would dominate the calculation time.

One way LQCD measurements can take advantage of the substantial incentives for large jobs is to group, or bundle, different computational tasks together.  
As LQCD measurements are often embarrassingly parallel, in that the measurements performed on different gauge configurations are independent of each other, these tasks can be bundled together quite simply.

\begin{figure}[t!]
    \centering
        \includegraphics[width=0.49\textwidth]{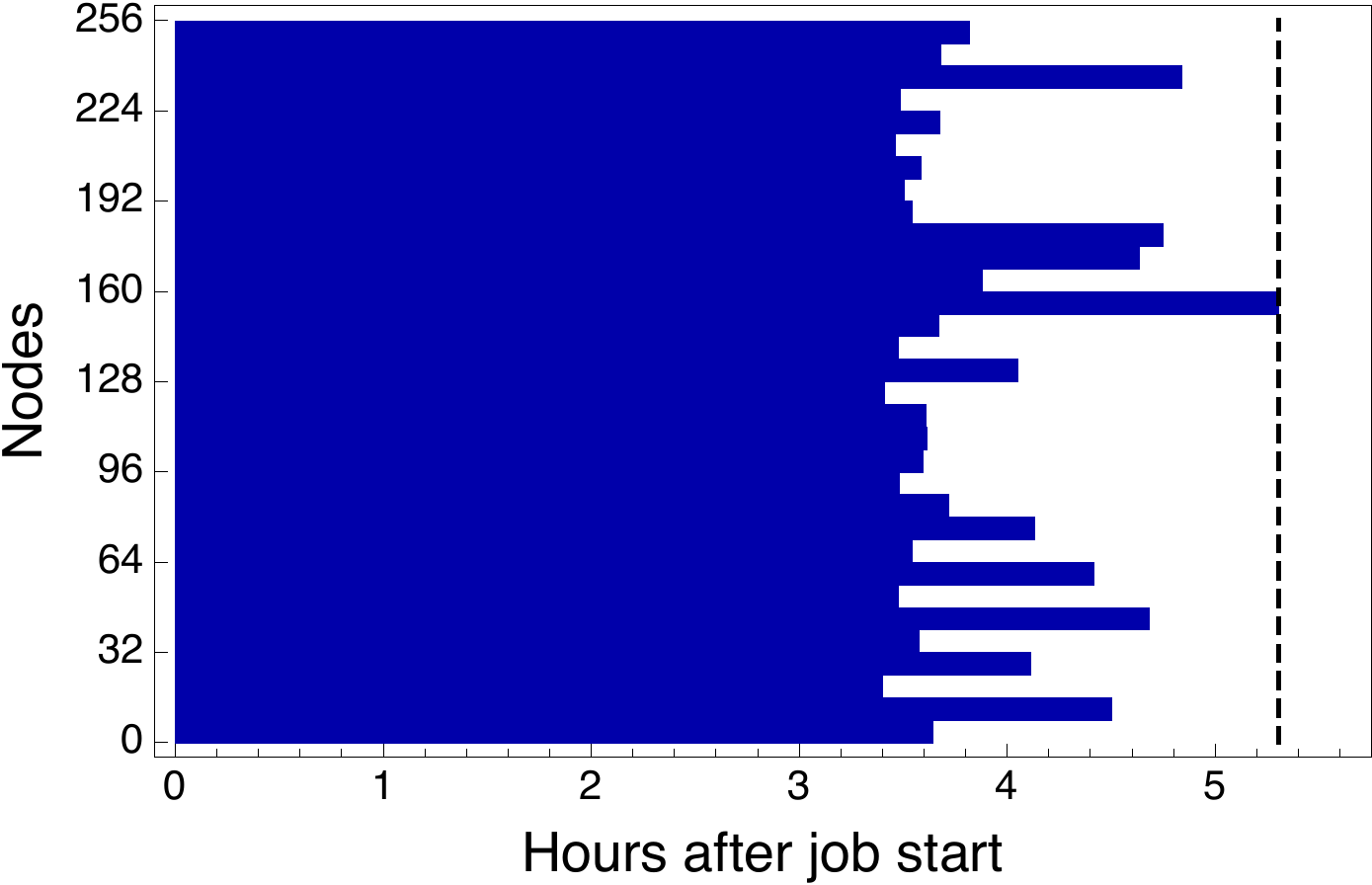}
        \includegraphics[width=0.49\textwidth]{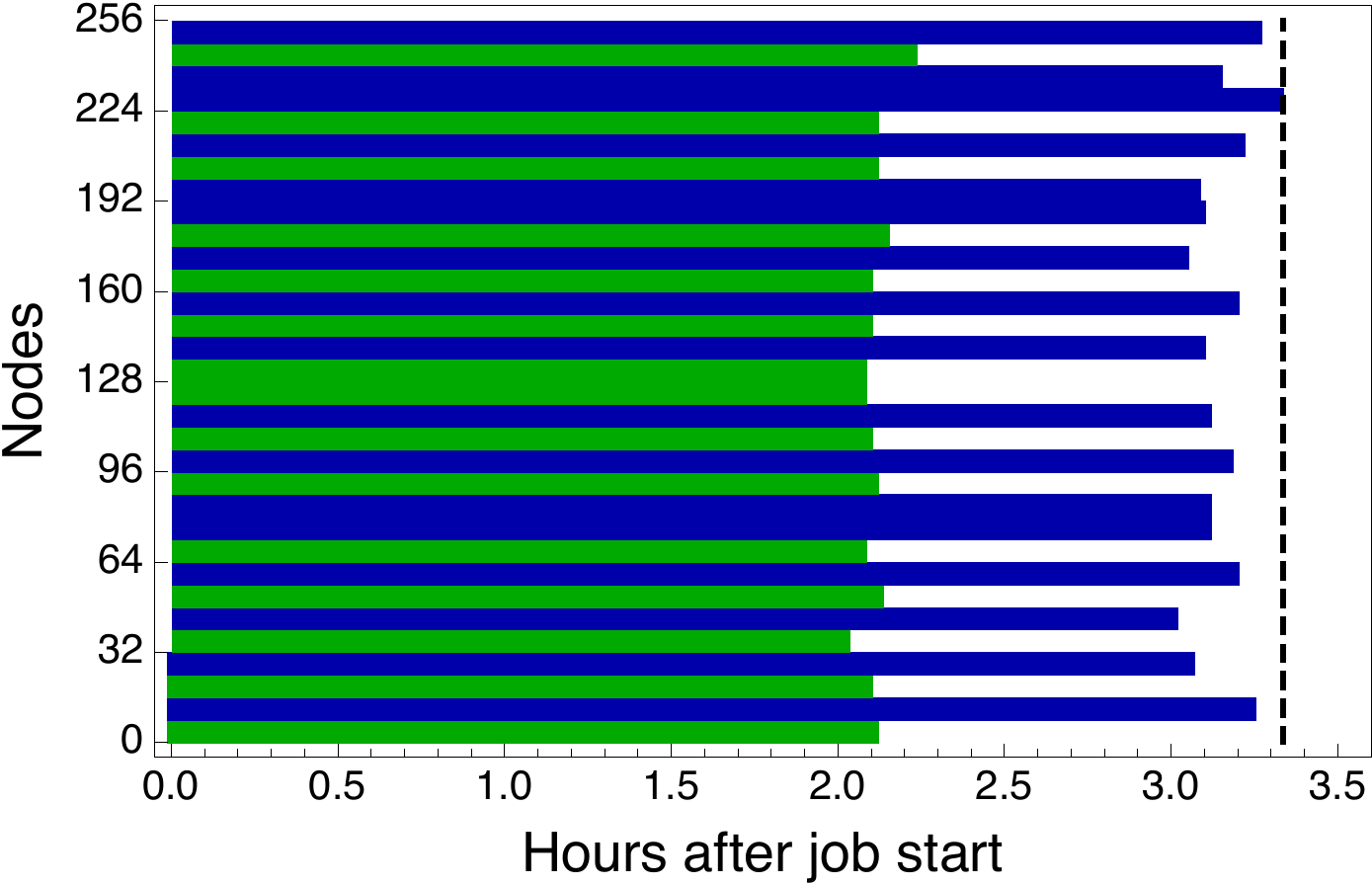}
    \caption{
Two timelines of multiple tasks na\"{i}vely bundled into a single 256-node job run on Titan.
The horizontal colored bars represent different computational tasks, and each color represents different types of tasks.
White space is wasted, idle time.
In the left panel we show how variation in run times for a single kind of task can vary, resulting in waste.
In the right panel we show how a heterogeneous mix of tasks can result in waste, even when the run time for each kind of task is relatively consistent.
}
    \label{fig:timelines-wasteful}
\end{figure}

One simple solution is na\"{i}ve bundling---finding similar computational tasks and starting them all at once with, for example, \wraprun\cite{wraprun}.
This bundling is na\"{i}ve, in the sense that it assumes different instances of similar tasks will finish at the same time.
However, as illustrated in the left panel of Fig.~\ref{fig:timelines-wasteful}, the wallclock time to complete different tasks of the same kind can vary dramatically.
The most common causes include differences in instance difficulty, on-node performance, and inter-node communication, but there are many others and they all lead to a substantial amount of wasted, idle resources.

As illustrated in the right panel of Fig.~\ref{fig:timelines-wasteful}, mixing different types of tasks can exacerbate the problem and lead to even more wasted resources, even if each task finishes roughly when expected.
Waste has the potential to undo the benefits of the incentives of large jobs.

\begin{figure}[bt!]
    \centering
        \includegraphics[width=0.49\textwidth]{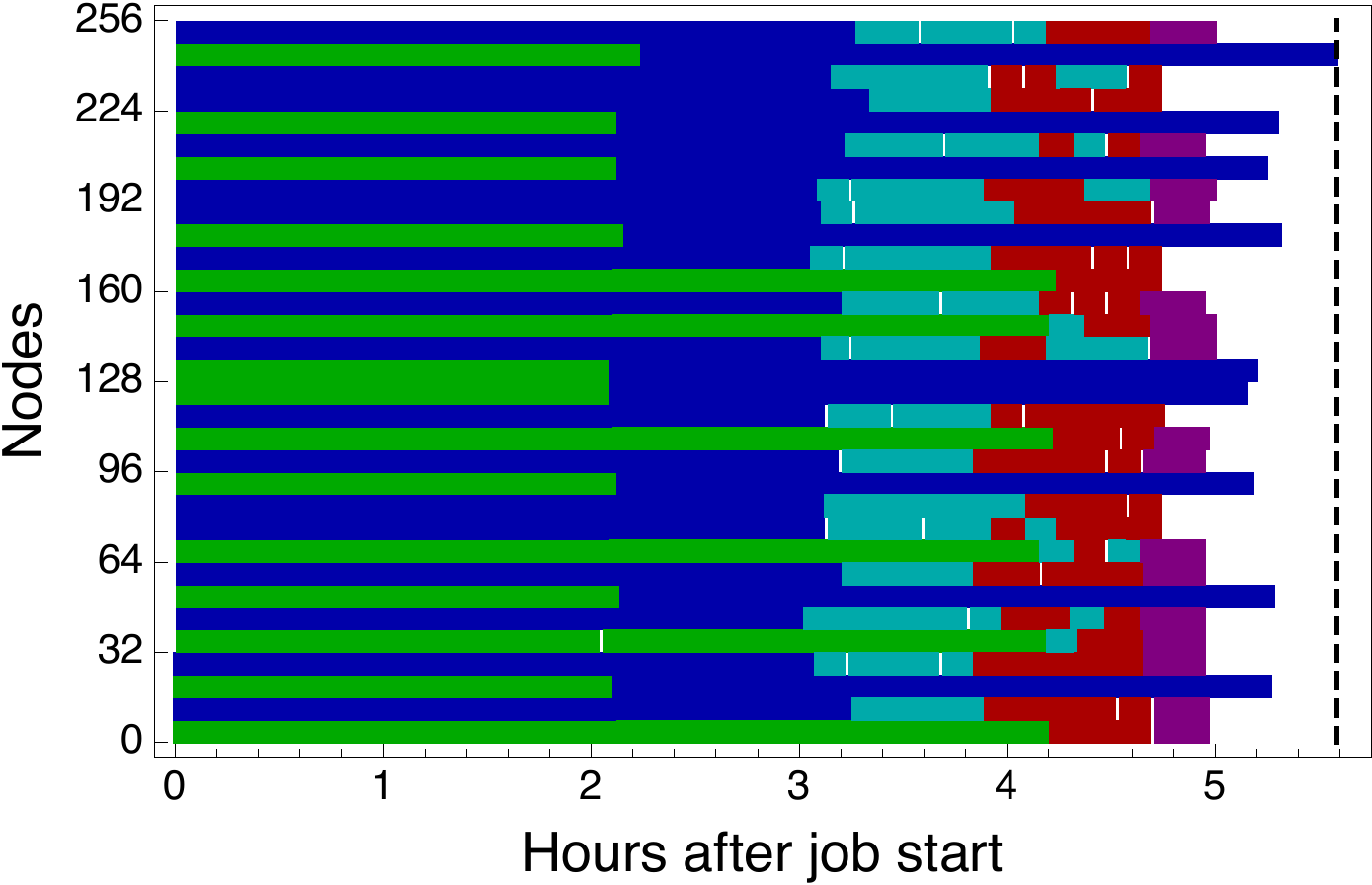}
        \includegraphics[width=0.49\textwidth]{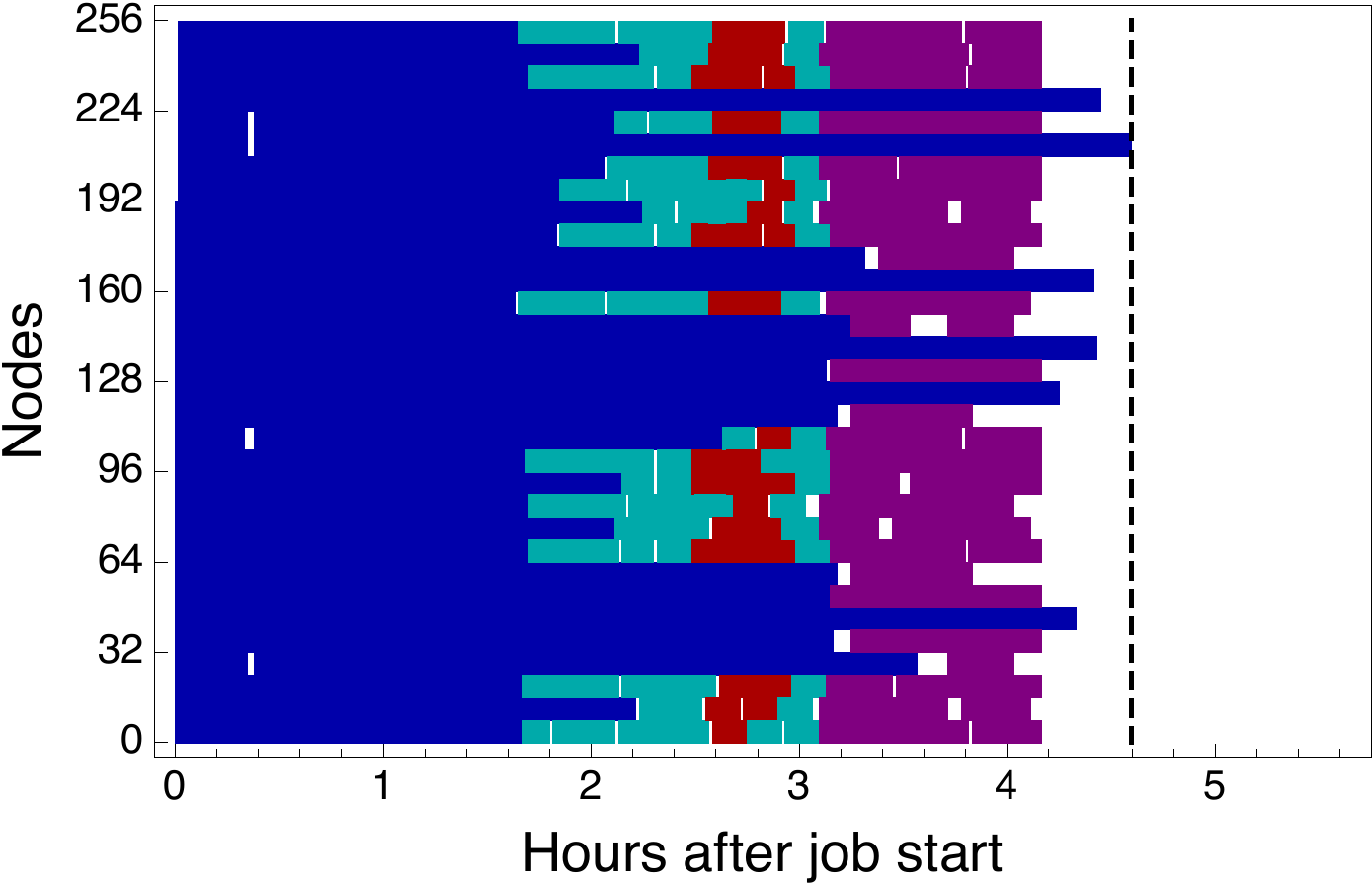}
    \caption{
Two timelines of multiple tasks bundled and backfilled into a single 256-node job run on Titan.
The horizontal colored bars represent different computational tasks, and each color represents different types of tasks.
White space is wasted, idle time.
}
    \label{fig:timelines-backfill}
\end{figure}

Backfilling computational tasks can markedly reduce the amount of wasted resources that can arise from na\"{i}ve bundling.
Backfilling is a widely adopted strategy in high-performance computing to minimize the amount of wasted resources by allowing light tasks to use idle resources, effectively at no additional cost.
This enables large bundles that can qualify for administrative incentives. 
By providing information about what resources a task requires---how many nodes, how much wallclock time on those nodes, etc.---one can achieve substantially less waste in large bundled jobs.
Figure~\ref{fig:timelines-backfill} shows two example jobs where a simple backfilling strategy, even without extremely accurate timing estimates, reduced the amount of waste that would have resulted from naive bundling.

To perform effective backfilling, the scheduling application must have information about the computational workload of each task.
The scheduler uses this information to solve an optimization problem to determine the optimal execution order at runtime.
Determining which tasks to compute at runtime rather than when the job is submitted to the batch scheduler's queue leads to other nice features.
For instance, multiple collaborators can create computational tasks that can be intermingled, and multiple collaborators can submit jobs that attack the same large set of computational tasks.
Additionally, a task can be changed after submission, and it is even possible to submit large jobs to the batch scheduler before deciding what tasks to schedule.

The dynamic selection of computational tasks at runtime also entails separating the job allocation request to the batch scheduler from the description of the computational work.
Traditionally, these are intermingled in a single script that is submitted to the batch scheduler.
Instead, to keep the job allocation independent from the tasks, independent descriptions are needed.
The job script submitted to the batch scheduler must be aware of the job's properties, such as wallclock time, number of nodes, etc., in order to configure the backfilling software.
The description of an individual task doesn't need these things---but it does need to know the resources required for the particular task, so that the backfiller can make a comparison between what resources are free and what are required to execute a task.

Here, we discuss \METAQ\cite{Berkowitz:2017vcp} and \mpijm\cite{mpijm}, two pieces of software designed to make bundling supercomputing tasks into large jobs easy.
Another pilot system\cite{turilli2015comprehensive}, \texttt{RADICAL-Pilot}\cite{merzky2015radical}, focuses on smaller, shorter jobs, and therefore optimizes for launch rate and number of concurrent tasks.
In contrast, LQCD calculations are often long-running and controlling communications efficiency and the ability to bundle tasks with different resource requirements are likely to yield the best performance.

\section{\METAQ}

\METAQ\footnote{Available at \url{https://www.github.com/evanberkowitz/metaq}, and licensed under GPLv3.0} is a simple implementation of software that can backfill computational tasks.
A manual is available on the arXiv\cite{Berkowitz:2017vcp}.
Implemented in \bash, it forms a proof-of-principle that a backfilling strategy can be used to waste fewer cycles.
However, as we will discuss, it has major drawbacks that has prompted the development of \mpijm, which will be discussed in the next section.
Nevertheless, \METAQ has been used successfully in production, and allowed us to intermingle the computations for Refs.~\cite{Berkowitz:2017gql,jason} with those for Ref.~\cite{Nicholson:2016byl} with little effort.
We use it on smaller clusters too, where bundling may not be necessary but the feature that collaborators can perform one another's work is useful.

\METAQ is designed to replicate the experience a user might already have grown used to when interacting with batch schedulers such as \SLURM\cite{SLURM}, \MOAB\cite{MOAB}, \TORQUE\cite{TORQUE}, or \PBS\cite{PBS}.
That is, the description of the computational tasks looks syntactically like a job script that might be submitted to one of those schedulers.
The user provides, using \texttt{\#METAQ} markup, information about the task and then is free to do any environmental setup and high-performance work (by invoking the batch scheduler's run command, such as \texttt{srun}, \texttt{aprun}, etc.).
This design ensures that executables remain unchanged.

In contrast to a real batch scheduler, there is no `submission'---the user simply puts the task description in file system directories.
The job scripts that are submitted to the batch scheduler look through these (user-configurable) directories for task scripts, compares the needed resources and wallclock time to what is currently available in the given allocation, and starts the task if possible.
As long as the file system permissions are set correctly, any collaborator may contribute or execute tasks.
Because the job scripts simply configure the backfilling, they are extremely uniform and simple to write.

Let us now discuss the drawbacks of \METAQ.
First, \METAQ trusts the user, in that if a task claims to only need a certain set of resources, those needs are not enforced. Therefore, if a task script lies, \METAQ's accounting of busy and idle resources will be incorrect, and the discrepancy may cause problems.

Second, \METAQ lacks the flexibility to execute the same task in different configurations.
For example, some tasks might run most efficiently on 16 nodes but can be done on 8 nodes if that's all that is available.
\METAQ does not have the ability to make these kinds of task-configuring decisions at runtime.

Third, the ability to assign work to an accelerator (say, a GPU) independently of assigning work to the CPUs is dependent on administrative policy.
For instance, this kind of overcommitting is not allowed at OLCF, but is possible on the Surface GPU cluster\cite{LLNL.surface} at LLNL.

Fourth, the lack of a `submission' utility means that the user must keep track of where the tasks belong.
In principle, a simple script may do, but none has been written as yet.

Fifth, because the tasks are represented by files on disk, iterating over the tasks can be slow and require examining the disk frequently.

Sixth, and most serious, the backfilling logic happens on the shared resources where the batch scheduler runs.
Because each run command (eg. \texttt{srun}, \texttt{aprun}, etc.) can be resource-intensive, and the backfiller itself launches monitoring processes for each task, it is possible to stress these resources.
Indeed, when scaling to very large jobs on Titan, we once crashed these service nodes and brought the machine down.
This incident resulted in a revision of the user guide, limiting the number of allowed simultaneous processes any job may use\cite{OLCF.processes}.

Many of these drawbacks are not necessarily problems but are consequences of \METAQ's implementation.
In the next section we will discuss \mpijm, software in preparation that is designed to address these drawbacks.

\section{\mpijm}

The success we had using \METAQ has inspired us to create \mpijm\cite{mpijm}, backfilling software that allows us to manage the computational resources allocated to a job much more finely.
Unlike \METAQ, which relied on the batch scheduler's run commands to move work to the compute nodes, \mpijm puts a daemon on each node at a job's start.
This daemon can monitor process IDs (PIDs), launch tasks as new processes, and can track resource consumption more reliably.

By design, \mpijm requires very little modification of binaries.
On compilation of an \mpijm-aware executable, one must include one file and link against one library.
Once compiled with \mpijm, a binary works whether or not it is launched under \mpijm management.
This is crucial for software maintenance and easy debugging.

Additionally, two handshake calls are required, just after \texttt{MPI\_Init} and \texttt{MPI\_Finalize}.
The daemon uses \texttt{MPI\_comm\_spawn}
%\footnote{IBM has promised that \texttt{MPI\_comm\_spawn} will be available in SpectrumMPI on next-generation machines.} 
to provide your executable with its own \texttt{MPI\_comm\_world}.
However, according to the \MPI specification, if any of these spawned processes die, the highest-level communicator will die as well.
Therefore, after the first handshake, the daemon disconnects from the executable so that a local error will only have local effects and will not cause the entire job and its other computational tasks to fail.
The second handshake is needed to send exit information from the disconnected processes back to the monitor daemon.

These minor modifications means that it is easy to integrate \mpijm into any software.
Indeed, we have already successfully integrated it with \QMP, the communications-wrapping layer of the USQCD software stack\footnote{An \mpijm-capable \QMP is available at \url{https://github.com/callat-qcd/qmp} on the \mpijm branch, and a pull request will be issued to the main \QMP repository when \mpijm is released.}.
By simply adding \texttt{--with-mpi-jm=/path/to/mpi\_jm} to the configuration of \QMP, any additional USQCD software libraries or user applications built on top of that \QMP installation will be \mpijm-aware.

As was found with \METAQ, inaccurate runtime estimates create waste.
One important source of inaccuracy when using a run command to launch tasks in the allocation of compute nodes is that as tasks of different sizes start and complete, the available nodes become fragmented.
Later launches may then be placed onto poorly-connected compute nodes.
To resolve this issue \mpijm partitions all the allocated nodes into user-configurable blocks.
To maximize communication performance, \mpijm uses information about the network topology to group neighboring nodes together\footnote{In future releases it might be possible determine configurations of fast blocks and storing that information, making it available machine-wide, or using more sophisticated methods for determining network bandwidth speeds.}.
Every task that can be considered must be able to use no more than the resources available in a block, but if it fits it will always get nodes with high performance connectivity, improving runtime and runtime predictability.
If a task needs 64 nodes but the blocks are 32 nodes each, that task will not run.

\begin{figure}[t!]
    \centering
        \includegraphics[width=0.49\textwidth]{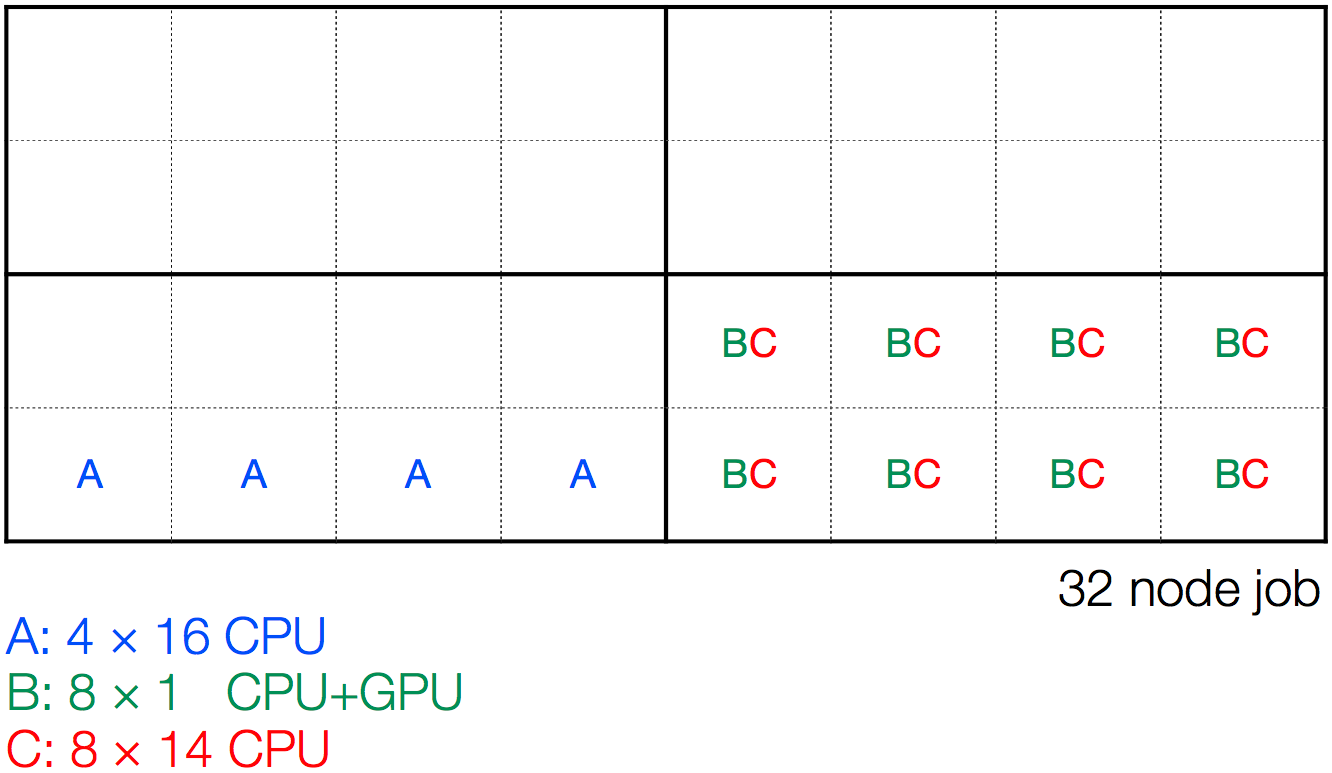}
    \caption{An example 32-node job on a hypothetical machine with 16 CPUs and 1 GPU per node, partitioned into 4 blocks of 8 nodes, being filled with three types of tasks (A, B, and C) which have different resource requirements.  Each node is indicated by a dotted box, while a block is indicated by a box of thick lines.  While A tasks require the whole node, B and C tasks can share nodes, B taking advantage of the GPUs, C only CPUs.}
    \label{fig:blocks}
\end{figure}

Each block shares a communicator with the overall scheduler, which is responsible for assigning tasks to different blocks.
The scheduler ensures that work isn't duplicated, and that work is only given to a block with the correct available resources.
Within a block, smaller or resource-saturating tasks may be launched, as schematically depicted in Fig.~\ref{fig:blocks}, where A tasks consume entire nodes but B and C tasks can peacefully share a node.

In the case where there are accelerators, \mpijm binds the appropriate CPUs to the GPUs based on configuration options and the hardware topology of the node.
Suppose a GPU-capable task just needs 1 CPU.
It may nevertheless make sense, to protect the CPU's caches, to think of it as requiring two CPUs, or at least to block other tasks' access.
On a node with 16 CPUs and a GPU, for example, it may be best for performance to share a node between a (1 CPU + 1 GPU) task and a 14-CPU task, leaving one CPU idle.
Moreover, it is conceivable that it's even better to give the CPU task only 12 CPUs, to limit competition for memory bandwidth, for example.
Tuning this kind of balance can be done on small jobs.
The resulting idle time is much less than otherwise achievable, and, as may increase overall performance and scientific output, should not be considered a loss, especially if one can execute different tasks on the same node simultaneously.

To create computational tasks we have implemented a \python interface to \mpijm.
This allows the user to create a python class for every task type.
That class can implement arbitrary pre- and post-execution steps and can implement a variety of configurations (for example, if a task can be run on different number of nodes).
One has access to the full power of \python, including its libraries, which makes it simple to handle logic surrounding task creation.

Currently, we describe tasks via \YAML\cite{YAML} dictionaries on disk, much as \METAQ's task scripts live on disk.
This means that currently \mpijm retains the fifth drawback discussed at the end of the previous section.
However, the examination and parsing of task scripts happens less frequently and is done at a much lower level (compared to \METAQ's \bash implementation).
In addition, the \mpijm scheduler doesn't run on the shared resources, but on the compute nodes.
This removes the possibility of taking down the service nodes and rendering the machine temporarily useless, and is substantially more neighbor-friendly.

Traversing file directories is not the only possibility for storing task descriptions---\python unlocks many possibilities.
One can imagine interfacing with a graph database or other task-generation, task-management, or automated data analysis suites\cite{taxi1,taxi2}.
Indeed, data reductions and analyses can themselves be run as tasks under \mpijm if it is logistically or economically beneficial.

Finally, because \mpijm relies only on the \MPI interface for communication, it can be used to manage resources across grids---for example, if they are connected only by \texttt{TCP/IP} (so long as \MPI works).
This has some interesting potential big-data applications, including for sharing computational resources across computing facilities (see, for example, Ref.~\cite{BigPanDA}) or eliminating idle cycles with tasks from users from afar.
One might even imagining issuing overlapping allocations which take advantage of different resources.

\section{Summary}

As we move towards the exascale era, the resources needed per problem instance for some computational tasks of scientific interest will stop growing.
To take full advantage of these machines, tasks can be bundled together.
\METAQ as a production-capable prototype, and \mpijm going forward, provide a path towards intelligent backfilling of computational tasks.

A dynamic, backfilling task scheduler not only allows one to scale up to large jobs to take advantage of administrative incentives, but also provides logistical benefits, such as ease of collaboration and a major reduction in the amount of wasted computational cycles.

\mpijm solves many of the issues with \METAQ, at the cost of very mild modification of one's binaries.
The needed modifications are already accessible to software built on top of the USQCD stack.
We hope to have a release candidate of \mpijm available soon, and anticipate distributing it widely with a liberal license.

\section*{Acknowledgements}

\METAQ was tested on \texttt{aztec}, \texttt{cab}, \texttt{surface}, and \texttt{vulcan} at LLNL through the Multiprogrammatic and Institutional Computing and Grand Challenge programs.
\METAQ was also tested and used in production on \texttt{titan} at Oak Ridge Leadership Computing Facility at the Oak Ridge National Laboratory, which is supported by the Office of Science of the U.S. Department of Energy under Contract No. DE-AC05-00OR22725, and \texttt{edison} and \texttt{cori} at NERSC, the National Energy Research Scientific Computing Center, a DOE Office of Science User Facility supported by the Office of Science of the U.S. Department of Energy under Contract No. DE-AC02-05CH11231.
This work was supported in part by the Office of Science, Department of Energy, Office of Advanced Scientific Computing Research through the CalLat SciDAC3 grant under Award Number KB0301052.
This work was supported in part by the DFG and the NSFC Sino-German CRC110.
This research used resources of the Oak Ridge
Leadership Computing Facility located at ORNL, which is supported by
the Office of Science of the Department of Energy under Contract No.
DE-AC05-00OR22725.

\clearpage
\bibliography{job-management}

%%%%%%%%%%%%%%%%%%%%%%%%%%%%%%%%%%%%%%%%%%%%%%%%%%%%%%%%%%%%%%%%%%%%%%%%%%%%%
\end{document}